\documentstyle[psfig,epsf]{l-aa}

\psrotatefirst

\begin{document}

\title{Constraining ${\Omega_{0}}$ from $X-ray$ properties of   
Clusters of Galaxies at high redshift}
\author{R. Sadat\inst{1,2}, A. Blanchard\inst{1} and J. Oukbir\inst{1,3}} 

\institute{ Observatoire Astronomique, 
            11, rue de l'Universit\'e, 67000 Strasbourg, France 
          \and
           C.R.A.A.G., BP 63, Bouzareah, Algiers, Algeria
          \and 
           D.S.R.I.,  Juliane Maries Vej 30, DK-2100, Copenhagen, Denmark
          }

\offprints{R. Sadat}

\date{Received \rule{2.0cm}{0.01cm} ; accepted \rule{2.0cm}{0.01cm} }

\maketitle
\markboth{Constraining ${\Omega_{0}}$ from $X-ray$ properties of   
Clusters of Galaxies at high redshift}{${\Omega_{0}}$ from $X-ray$ Clusters  at high redshift}
 
\begin{abstract} 
Properties of high redshift clusters are a fundamental source of 
information for cosmology. It has been shown by Oukbir and Blanchard 
(1997) that 
the combined knowledge of the redshift distribution of X-ray clusters of 
galaxies and the  
luminosity-temperature correlation, $L_X-T_X$, provides a powerful test of 
the mean density of the  Universe. In this paper, we address the question 
of the possible evolution of this relation from an observational point 
of view and its cosmological 
significance.
We introduce a new indicator in order to measure the evolution of the X-ray 
luminosity-temperature relation  
with redshift and take  
advantage of the recent availability of temperature information 
for a significant number of high and intermediate redshift X-ray  
clusters of galaxies.  
From our analysis, we find a slightly positive 
evolution in the $L_X-T_X$ relation. 
This implies 
a high value of the density 
parameter of $0.85\pm0.2 $. However, because the selection 
of clusters included in
our sample  is unknown, this can be considered only as a tentative 
result. A well-controlled X-ray selected survey would provide a 
more robust answer. XMM will 
be ideal for such a program.
\end{abstract}  
\section {Introduction} 
 
Clusters of galaxies are ideal tools for cosmology since they are the largest  ``virialized" structures in the Universe. They are strong X-ray emitters, 
and as such they provide  
useful information on the evolution and formation of structures in the 
universe. 
 In particular, cluster evolution can be  
inferred from the study of X-ray properties of distant clusters. 
There has been 
some debates on the existence and the nature of the evolution of X-ray 
clusters. Oukbir, Blanchard \& 
Bartlett (1997, OBB97 hereafter) established a completely self-consistent modeling of 
X-ray clusters, and concluded
that all the available data can be reproduced with a much 
lower rate of evolution  than inferred from the EMSS survey 
(Edge et al., 1990; Henry and Arnaud, 1991, HA91 hereafter). From recent 
analyses, it seems more and more clear that the X-ray luminosity function 
evolves 
(Collins et al., 1997;
Nichol et al., 1997; Ebeling et al., 1997), 
but at a lower rate than
has been previously reported.

The Press-Schechter formalism (1974) has been extensively used in order 
to reproduce the global properties
of X-ray clusters (Henry \& Arnaud, 1991; Blanchard et al, 1992;  
Bartlett \& Silk, 1993; 
Colafrancesco \& 
Vittorio, 1994; De Luca et al, 1995 ; Viana \& Liddle, 1996; Eke et al, 1996; 
Kitayama \&  Suto, 1997; Bahcall et al., 1997; Mathiesen \& Evrard, 1997 among others) 
as this formalism seems to describe  
accurately the clusters distribution of mass, $m$, and redshift, $z$.
This has also been applied to Sunyaev-Zeldovich  number counts 
(Barbosa et al, 1996). 
Oukbir \& Blanchard (1992) showed that the existence of high temperature 
clusters at high redshift is more likely in open universes than in a universe
with the  critical density. 
Oukbir \& Blanchard 
(1997, OB97 hereafter) showed that the relative evolution of cluster  
abundance 
depends only on the growth rate of structure, which depends on the 
cosmological parameters of universe, but not on the spectrum of the 
primordial fluctuations. On this basis, they  have  
proposed a new test for constraining (${\Omega_{0}}$, ${\Lambda}$): 
the redshift  
evolution of the X-ray cluster temperature distribution function. 

Such a test requiring  the 
knowledge of  the evolution  of the temperature distribution 
function is difficult to apply. Henry (1997) has shown a first application, 
while Donahue et al. (1997) have presented similar 
considerations. Carlberg et al. (1997), C97 hereafter, present 
a tentative application of 
this test on the basis of velocity dispersion. Bahcall et al. (1997) 
also present an application of this test.
 However, by fitting the evolution of the 
luminosity-temperature relation in 
order to match 
the redshift distribution of EMSS clusters, OB97 showed 
that knowledge of the evolution of  
this quantity 
allows an equivalent determination of  ${\Omega_{0}}$ in an efficient way. 
Few attempts have been made to measure the evolution of 
the $L_{X}-T_{X}$ relation  
and it has not yet been applied to constrain cosmological parameters. 
In this paper we attempt for the first time a comparison of the  
theoretical predictions of the $L_{X}-T_{X}$ relation with available 
data we have gleaned from  
the literature. In section 2, we briefly discuss the foundations of this test.
In section 3 we discuss the set of clusters we have used to perform 
the test. In section 4, 
we describe the method we have used to estimate the possible evolution. 
    
\section{Clusters properties and the mean density of the universe} 
 
The Press-Schechter (1974) formalism seems to give an accurate 
determination of the mass  
function. The reason for this has been a matter of debate, but 
it has allowed investigation of the  
non-linear evolution of structure formation in much detail. This 
has been widely used to  put constraints on 
cosmological parameters, such as the amplitude and the shape  
of the power spectrum on galaxy clusters scales by comparison with 
observations. However, 
cluster masses are 
not directly measured,  
therefore it is necessary to establish  relations between the mass 
and ``observables" such 
as the X-ray luminosity or  
the X-ray temperature. As the luminosity of a cluster is difficult 
to relate to its virial 
mass from theoretical 
arguments, it has been argued by some authors that the temperature 
distribution function is more adequate for  
a fruitful comparison, although Balland \& Blanchard (1997) showed that 
hydrostatic equilibrium does not provide a one-to-one correspondence
between mass and temperature. This relation should therefore be taken 
from the numerical simulations where  gas dynamics are taken into account
(Evrard, Metzler \& Navarro 1996). 
 The consequence for  cosmology of the observed X-ray luminosity and  
X-ray temperature distribution functions has been investigated in 
recent years.  OBB97 have shown that a comprehensive description 
can be constructed, in a consistent way, provided that the relation 
between luminosity and temperature is specified (from the observations). 
Such a scheme 
has been used by OB97 to investigate the cosmological implication of 
X-ray clusters in an open cosmology  
and to establish a self-consistent modeling in such a context. 
They have shown that although the properties 
of X-ray clusters at redshift zero can be well reproduced in 
an open model, the redshift evolution is significantly 
different: this is due to the fact that in open model universes the growth 
rate of fluctuations is lower than in a Einstein--de Sitter universe. 
Therefore, at high redshift, a higher 
number density is expected than in an $\Omega_0 = 1$ universe. 
Specifically, they prove (see their section 4) that  
the redshift evolution of the number of clusters of a given mass, 
or equivalently of a given apparent temperature, is almost independent of 
the spectrum of the primordial fluctuations, but that it 
depends on the mean density of the universe (and on others 
cosmological parameters of the universe). The evolution of the mass function 
therefore allows one to measure 
the mean density of the universe (and the cosmological constant),  providing
 a new cosmological test, based on the dynamics of the universe as a whole. 
\begin{table}[ht]
\begin{center}
\begin{tabular}{lllllcllclllcr}
  \hline
Cluster & z &$T_{X}$ & $L_{bol}$ \\
        &   & ${\sf keV}$ & $10 ^{44} {\sf erg/s}$& Ref.\\
\\
\hline
\\
Virgo	   & 	0.0038  & 2.34$^{+0.02}_{-0.02}$  &0.68 &{\sl 28} \\
Centaurus  &	0.01    & 3.9$^{+0.2}_{-0.20}$  &1.22   &{\sl 1}   \\
A1060	   &	0.0114  & 3.1$^{+0.3}_{-0.5}$  & 0.658&{\sl 22}  \\
A262	   &	0.0164  & 2.4$^{+0.3}_{-2.20}$  & 0.85 &{\sl 1} \\
AWM7	   &	0.0176  & 4.0$^{+0.3}_{-0.20}$  & 2.76 &{\sl 1}\\
A426       &    0.0183  & 6.3$^{+0.3}_{-0.3}$  & 23.1 &{\sl 1}\\
A539 	   &	0.0205  & 3.0$^{+0.8}_{-0.6}$ & 0.64 &{\sl 1} \\
A1367	   &	0.0215  & 3.5$^{+0.18}_{-0.18}$  & 2.2  &{\sl 1} \\ 
3C 129	   &	0.0218  & 6.2$^{+0.8}_{-0.6}$  & 3.7 &{\sl 1}  \\ 
A1656	   &	0.0232  & 8.11$^{+0.07}_{-0.07}$  & 17.2 &{\sl 1} \\
Ophicius   &	0.028   & 9.8$^{+0.7}_{-0.3}$  & 27.4  &{\sl 22}\\
A2199	   &	0.0299  & 4.5$^{+0.3}_{-0.2}$  & 6.4  &{\sl 1} \\
A496	   &	0.032   & 3.91$^{+0.06}_{-0.06}$ & 7.16 &{\sl 1} \\
A576	   &	0.0381  & 4.3$^{+0.5}_{-0.4}$  & 2.94 &{\sl 1} \\
A3558	   &	0.048   & 5.5$^{+0.3}_{-0.2}$ & 10.  &{\sl 1,8}\\ 
Triangulum &	0.051   &10.3$^{+0.8}_{-0.8}$  & 30. &{\sl 12}\\ 
A85	   &	0.052   & 6.2$^{+0.4}_{-0.5}$  & 15.8&{\sl 1} \\
A3667	   &	0.053   & 6.5$^{+0.8}_{-0.99}$& 21.4&{\sl 15}\\
A754	   &	0.0534  & 8.5$^{+0.82}_{-0.82}$  & 29.4  &{\sl 26} \\ 
A2319	   &	0.0564  & 10.0$^{+0.7}_{-0.7}$  & 37.   &{\sl 11} \\ 
A2256	   &	0.0601  & 7.51$^{+0.19}_{-0.19}$  &  24.3&{\sl 1} \\
A1795	   &	0.0616  & 6.7$^{+0.66}_{-0.66}$ & $23.75^{*}$ &{\sl16}\\
A399   	   &	0.07    & 7.0$^{+1.0}_{-1.0}$  &  14.2&{\sl 5}\\
A644	   &	0.0704  & 6.6$^{+0.17}_{-0.17}$   &  23.5&{\sl 1}\\
A401	   &	0.0748  & 8.0$^{+1.0}_{-1.0}$  &  30.8&{\sl 5}\\
A2142	   &	0.0899  & 9.0$^{+0.2}_{-0.2}$ &  $56^{*}$ &{\sl 30} \\
    \hline
        \end{tabular}
 \caption{\small Temperatures and bolometric luminosities for a sample of 
low-redshift clusters. Asterisks indicate clusters considered as having a 
strong 
cooling flow. The numbers in column 6 indicate the references from which 
the luminosity and the temperature, respectively, are taken ($\Omega_0 = 1$ 
and $H_{0}=50$ {{\rm km/s}/Mpc}). The quoted error bars are given at the 90\% confidence level.} 
         \label{Table 1a}
\end{center}
            \end{table}
 
\subsection{Measuring $\Omega_0$ with clusters} 
 
OB97 and OBB97 have used the observed correlation $L_{X}-T_{X}$ 
to construct a self-consistent modeling of X-ray clusters. The 
relation between luminosity and temperature they used 
is: 
\begin{equation} 
L_{bol }= L_1 T_{\rm keV}^{\alpha} 
\end{equation} 
 with $ L_1 = 0.049 \; {10^{44} {\rm erg/s/cm}^2}$ and $\alpha = 3$. A
 more recent 
analysis (Arnaud \& Evrard, 1997) showed that the  $L_{X}-T_{X}$ does 
have a moderate 
intrinsic dispersion when cooling flow clusters are removed,
 with $ L_1 = 0.067 \; 10^{44} {\rm erg/s}$ and 
$\alpha = 2.89$. By comparing to the EMSS cluster redshift distribution, 
OB97 have shown that in the absence of evolution of the $L_{X} - T_{X}$ relation,
 open models   
with ${\Omega_{0}} {\sim} 0.2 $  
predict much more clusters than observed while $\Omega_{0}= 1$ model 
fits the data reasonably well,  although 
the abundance of high redshift clusters was not very well reproduced.

They 
have investigated the possibility of  
evolution by allowing the relation $L_{X} - T_{X}$ to change 
with redshift according to the following 
 form : 
  \begin{equation} 
L_{bol} = L_1 (1 + z)^{\beta} T_{\rm keV}^{\alpha} 
\label{LTevol}
\end{equation} 
where 
${\beta}$ is a free parameter  
which can be derived by fitting the redshift distribution of
the cluster EMSS survey. They found that 
${\beta}=1$ for a flat universe (${\Omega_0}=1$), corresponding
 to positive evolution, while a significant 
negative  evolution, corresponding to  
${\beta} = - 2.3$ was required for an open universe (${\Omega_0} \sim 0.2$).
 Following a 
similar approach and  by fitting the ROSAT number counts,  
Kitayama \& Suto (1997) provided constraints on the parameters of 
the model and Mathiesen \& Evrard (1997) reached essentially 
identical conclusions. Kitayama et al. (1997) has extended the predictions to the SZ counts. The self-similar model predicts $\alpha = 2.0$ and 
$\beta = 1.5$, but $\alpha = 2.$ and is clearly ruled out by observations, and 
physical processes specific to the baryonic content have to be advocated.
For instance, HA91 assumed an isentropic model. Bower (1997) has 
recently re-examined this question in more detail.  
   
\section {The high-redshift cluster sample} 

\begin{table*}[ht]
   \begin{tabular}{lllll}
      \hline
Cluster & z &$T_{X}$ & $L_{bol}$ \\
        &   & ${\sf keV}$ & $10 ^{44} {\sl erg/s}$& Ref.\\
\hline

A2204	   &	0.153   & 8.5$^{+0.4}_{-0.45}$    &  76. &{\sl 2}\\
A3888	   &	0.168   & 7.9$^{+0.3}_{-1.0}$  &  33. &{\sl 17}\\
A1204      &	0.17    & 3.6$^{0.13}_{-0.13}$&  14. &{\sl 18}\\
A586	   &	0.171   & 6.8$^{+0.7}_{-0.67}$  &  17.96&{\sl 3}\\
RXJ 1340+4018&  0.171   & 0.92$^{+0.13}_{-0.13}$ &   0.45&{\sl 27}\\
A2218	   &	0.175   & 6.72$^{+0.83}_{-0.83}$  &  19.3&{\sl 1}\\
A1689	   &	0.181   & 8.7$^{+0.51}_{-0.49}$  &  47.9 &{\sl 3}\\
A665	   &	0.182   & 8.9$^{+0.62}_{-0.61}$   &  28. &{\sl 3}\\   
A1763	   &	0.187   & 9.0$^{+1.02}_{-0.84}$  &  35.5&{\sl 6,2}\\
A1246	   &	0.187   & 6.3$^{+0.54}_{-0.51}$  &  19.5&{\sl 2}\\
SC 2059-25  &   0.188   & 7.0$^{+6.9}_{-2.2}$  &23.15 &{\sl 9}\\
MS0839.8   &	0.194   & 3.8$^{+0.4}_{-0.31}$  &   7.4&{\sl 3} \\
A2507      &    0.196   & 9.4$^{+2.7}_{-1.9}$  &  40. &{\sl 1}\\
MS0440	   &	0.1965  & 5.6$^{+0.80}_{-0.60}$  &   6.1&{\sl 31,19} \\
A2163	   &	0.201   & 14.6$^{+0.9}_{-0.80}$  & 143.  &{\sl 13} \\
A520(MS0451+02)&0.201   & 8.6$^{+0.93}_{-0.90}$   &  17. &{\sl 4,2} \\
A963	   &	0.206   & 6.76$^{+0.44}_{-0.49}$  &  21.&{\sl 2}\\
A1851	   &	0.2143  & 5.04$^{+0.8}_{-0.68}$  &   6.23&{\sl 3} \\
A773	   &	0.217   & 9.6$^{+1.03}_{-0.90}$  &  30. &{\sl 6,2} \\
A1704	   &	0.219   & 4.5$^{+0.56}_{-0.34}$  &  13. &{\sl 2}\\
A1895	   &	0.225   & 6.7$^{+1.38}_{-1.05}$  &  10.5&{\sl 3}\\
A2390	   &	0.228   & 8.9$^{+0.97}_{-0.77}$  &  54. &{\sl 2} \\
A2219	   &	0.23    &11.8$^{+1.26}_{-0.74}$  &  57. &{\sl 6,2}\\
MS1305.4   &	0.241   & 2.98$^{+0.52}_{-0.41}$  &   2.86     &{\sl 2} \\ 
A1835	   &	0.252   & 9.1$^{+2.10}_{-1.30}$  &  $80.^{*}$ &{\sl 7}\\
Zw 7160    &	0.258   & 5.2$^{+2.2}_{-0.70}$  &  $30.^{*}$ &{\sl 7}\\
A348       &	0.274   & 4.85$^{+0.6}_{-0.6}$  &   3.6  &{\sl 36}\\  
A33        &	0.28    & 4.06$^{+0.5}_{-0.5}$  &   2.4  &{\sl 36}\\ 
A483	   &	0.28    & 8.7$^{+3.3}_{-2.2}$   &  48.9 &{\sl 1}\\
\hline
\end{tabular}
\hspace*{2mm}
   \begin{tabular}{lllll}
      \hline
Cluster & z &$T_{X}$ & $L_{bol}$ \\
        &   & ${\sf keV}$ & $10 ^{44} {\sl erg/s}$& Ref.\\
\hline
A1758N	   &	0.28    &10.2$^{+2.3}_{-1.7}$    & 29.&{\sl 6,2} \\
Zw3146	   &	0.291   & 6.2$^{+1.2.}_{-0.7}$  &  $64.^{*}$&{\sl 7}\\
A1722	   &	0.301   & 5.87$^{+0.51}_{-0.41}$  &  21.&{\sl 2}\\ 
MS1147.3   &    0.303   & 5.5$^{+1.32}_{-1.00}$ &  6.6& {\sl 32}\\
MS1008	   &	0.306  & 7.9$^{+2.00}_{-1.65}$  &  16.32&{\sl 32}\\  
AC118	   &	0.308   &9.33$^{+1.09}_{-0.83}$  &  48. &{\sl 3}\\  
MS1241.5   &    0.312   & 6.2$^{+3.00}_{-2.15}$  &   6.4& {\sl 32}\\
MS0811.6   &    0.312   & 4.6$^{+1.50}_{-1.00}$ &   4.76& {\sl 32}\\
MS2137     &    0.313   &4.7$^{+0.5}_{-0.3}$  & 34.19&{\sl 32}\\
A1995	   &	0.318   & 9.44$^{+2.18}_{-1.54}$  &  22.5&{\sl 3}\\
MS0353	   &	0.32    & 6.2$^{+1.65}_{-1.32}$  &  12.85&{\sl 32} \\
MS1426.4   &    0.32    & 5.5$^{+1.82}_{-1.15}$  &   8.85& {\sl 32}\\
MS1224.7   &    0.326   & 4.3$^{+1.15}_{-1.00}$ &   7.46 &{\sl 32} \\
MS1358	   &	0.329  & 6.6$^{+0.82}_{-0.82}$   &  18.94&{\sl 32}\\   
A959	   &	0.353   & 6.47$^{+1.15}_{-1.01}$   & 16.   &{\sl 3}\\
MS1512     &    0.3726  & 3.8$^{+0.66}_{-0.50}$  &  8.212 &{\sl 32}\\
A370	   &	0.373   & 6.39$^{+1.02}_{-0.81}$  &  21.7&{\sl 3} \\
Cl 0939+47 & 	0.41  &  2.9$^{+1.3}_{-0.8}$ &  11.&{\sl 34}\\
Cl 09104+4109	&0.442  &  11.4$^{+3.2}_{-3.2}$ &   75.&{\sl 20}\\
RXJ1347.5-1145& 0.451  &  9.3$^{+1.1}_{-1.0}$ &  $210.^{*}$&{\sl 23}\\
A851	   &	0.451   & 6.7$^{+2.7}_{-1.70}$    &  15.&{\sl 2}\\
3C295	   &    0.46    & 7.13$^{+2.06}_{-1.35}$  &  24.&{\sl 2}\\
MS0451-03  &	0.5392  &10.4$^{+1.60}_{-1.30}$  &  46.8&{\sl 24}\\
RXJ 0018.8+160&0.544   & 1.6$^{+0.7}_{-0.4}$ 0.6  &   3.2& {\sl 29}\\
CL0016+16  &	0.5466  & 7.55$^{1.188}_{-0.957}$  &  55.&{\sl 10}\\
RXJ1716+67 &    0.813   & 6.7$^{+2.0}_{-2.0}$   &  17.72&{\sl 33}\\
MS1054.5-0321&	0.826   &14.7$^{+4.6}_{-3.5}$  &  42.&{\sl 25}\\
AXJ2019    &	1.0     &8.6$^{+6.9}_{-4.9}$  &  19    &{\sl 35}\\
 & & & & \\
       \hline
          \end{tabular}
		\caption{\small X-ray temperatures and bolometric luminosities
 for high redshift clusters.
          \label{Table 1b} }

${\bf Ref.}$ 
{\sl (1)} David et al., 1993;
{\sl (2)} Mushotzky \& Scharf, 1997;  
{\sl (3)} Tsuru et al., 1996;
{\sl (4)} Nichol et al., 1997; 
{\sl (5)} Fujita et al. 1997; 
{\sl (6)} Ebeling et al. 1995; 
{\sl (7)} Allen et al. 1995;  
{\sl (8)} Markevitch \& Vikhlinin, 1997;  
{\sl (9)} Arnaud et al., 1991;  
{\sl (10)} Hughes \& Birkinshaw, 1997;
{\sl (11)} Markevitch, 1996;  
{\sl (12)} Markevitch et al., 1996;  
{\sl (13)} Elbaz et al., 1996;  
{\sl (14)} Tamura et al., 1996;  
{\sl (15)} Knopp et al., 1996;  
{\sl (16)} Briel \& Henry, 1996;  
{\sl (17)} White \& Fabian, 1996;  
{\sl (18)} Matsuura et al., 1996 ;
{\sl (19)} Gioia private communication ;
{\sl (20)} Hall et al., 1996; 
{\sl (21)} Hamana et al., 1997; 
{\sl (22)} Matsuzawa,et al., 1996 ;
{\sl (23)} Schindler et al., 1996; 
{\sl (24)} Donahue, 1996 ;
{\sl (25)} Donahue, et al., 1997 ;
{\sl (26)} Henriksen \& Markevitch, 1996;
{\sl (27)} T.J, Ponman et al., 1994;
{\sl (28)} Arnaud \& Evrard, 1997;
{\sl (29)} Connolly et al., 1996;
{\sl (30)} White et al., 1994;
{\sl (31)} Gioia \& Luppino, 1994; 
{\sl (32)} Henry 1997; 
{\sl (33)} Henry \& Gioia 1997;
{\sl (34)} Schindler, 1997;  
{\sl (35)} Hattori et al., 1997;  
{\sl (36)} Colafrancesco, 1997;
\end{table*} 

In order to perform this new cosmological test, we have collected from the 
literature all the 
information on temperature measurements of X-ray clusters.
Although the number of high redshift clusters ($z > 0.4$) with an estimation 
of the X-ray 
temperature remains small, ASCA observations of clusters of relatively high 
redshift ($z \approx 0.3$) has begin to appear in the literature.  
We have tried to compile all the existing X-ray clusters with redshift 
greater than 0.15  for which the temperature has been  
measured with various X-ray satellites, mostly ASCA. 
This represents 57 clusters whose properties are summarized in Table 2.\\  
In order to address the question of the possible evolution 
of clusters, we have decided to include some clusters for 
which a reliable mass estimate 
is available and derived the corresponding temperature. This has led us to 
include the CNOC survey of clusters by C97, 
as it  contains accurate velocity dispersion measurements, and 
the sample of Smail  et 
al. (1997) (hereafter S97), who have used deep HST images to study  
weak shear of background galaxies by distant clusters of galaxies to estimate  
their masses. From these two samples, only clusters for which 
temperature information  was not 
available were retained in the analysis. The final compilation 
contains a total number of 57 clusters at  
high redshift ($z > 0.15$), 26 being at redshift greater than  $  0.3$.
The most distant   
redshift clusters of the sample are MS1054.5-0321 at 
z = 0.83 which has been measured by  Donahue et al. (1997) 
and AXJ2019 at z = 1 by Hattori et al. (1997).
An additional list of clusters at low redshift, for which accurate 
temperature information was available, 
has been added to the sample for the purpose of our analysis, 
but we do not attempt to be complete. This set is used as a template (Table 1). \\

Ideally, in order to perform the cosmological test, it would be 
better to estimate the evolution of  
$L_X-T_X$ from the EMSS sample itself. However, too few measurements 
exist up to now.  
Still, our sample contains a number of high redshift clusters which 
is comparable  
to that of the EMSS sample in the various redshift ranges. 
For comparison, the EMSS survey  
(Gioia \& Luppino 1984) has the same number of clusters at $z > 0.3$  
but a total number of $z > 0.15$ of only 49 (Figure 1). This 
is important as it does mean
that we are testing the evolution of the $L_X-T_X$ relation 
over the same redshift range
as OB97 have investigated, with a similar number of clusters involved.
\begin{figure}[h] 
\epsfxsize=9cm 
\centerline{\epsfbox{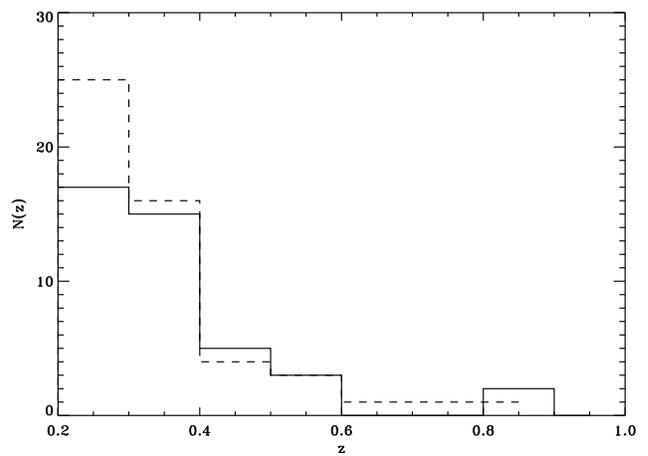}} 
\caption{\small \sl The redshift distribution in our sample (full line)
compared with the one in the EMSS cluster 
 survey (hashed line). } 
\label{Fig. 1} 
\end{figure} 

\subsection{Fluxes} 
 
As  the available data come from different satellites, the published  
luminosities are given in various energy bands.  
Therefore, when the bolometric luminosity was not available, it has been 
estimated by using a   
Raymond--Smith code, taking an  
average metallicity of 0.33 when its value was not available (see Tables 1, 2). 
Flux calibration is a serious worry when data from different 
satellites are used.
Arnaud \& Evrard (1997) have shown that GINGA and ROSAT fluxes agree very well,
while an offset in the calibration of EXOSAT is suspected. 
In some cases, several flux estimates are given, which disagree 
from time to time.  
Our fluxes were taken from different authors: 
David et al. (1991),  Mushotzky  \& Scharf  (1997) 
(hereafter {M\&S97}),   
Ebeling (1996) from ROSAT PSPC observations and Nichol {\it et al.} (1997)  
from HRI observations. In general, we have preferred ROSAT 
 measurements
when available. 
 
\subsection{Cooling flows} 
 
Cooling flow clusters present a central enhancement in their 
luminosity, in a region were the  
cooling time of the gas is shorter than the Hubble time. The 
inclusion of such clusters in our analysis 
is problematic : as they are more luminous than ``normal clusters", 
they might introduce a bias. 
However, as the EMSS is flux selected, there is no reason 
to assume that cooling flow clusters are not present in the 
EMSS sample as well. Their point-like nature could even 
represent a bias favoring their presence in the EMSS, as the high 
background makes the detection of extended source more dificult.
Still, it is also conceivable that cooling flow clusters are overabundant 
in the general compilation compared to the EMSS sample.
We have therefore applied a correction to 
clusters for which the presence of a cooling flow was known: only 
the flux outside the cooling flow was taken 
into account. Such cooling flow clusters are flagged by an asterisk. 
In practice, this correction is never larger than 50\%.  
This has been applied only to a few number of clusters. \\
\begin{table}[t] 
\begin{center}
\begin{tabular}{llrrr}
\hline

Cluster & z &  ${\sigma}$~~ &$T_{est}^{v}$&$T_{X}$ \\
        &   &  ${\sf km/s}$& ${\sf keV}$ &${\sf keV}$\\   
\hline
 
A2390  &0.228  &  1104 &   9.95  &     $8.9^{a}$ \\ 
MS0440 &0.1965 &  606 &   3.00  &     $5.6^{a}$ \\      
MS1008 &0.306  &  1054 &   9.06  &     $7.7^{a}$ \\   
MS1358 &0.329  &   934 &   7.12  &     $6.5^{a}$ \\       
MS1512 &0.373  &  690 &   3.88  &     $4.2^{a}$ \\  
MS0451+02 &0.2011&1031 &   8.70   &     $8.6^{a}$  \\  
MS0451-03 &0.5392&1371 &  15.30   &    $10.4^{b}$  \\  
MS0839 &0.1928 &   756 &   4.66  &     $3.8^{c}$ \\  
MS1455 &0.2568 &1133   &  10.50   &     $5.0^{a}$  \\  
MS1224 &0.3255 & 802   &   5.25  &     $4.3^{d}$  \\  
\hline
\end{tabular} 
\caption {\small Comparison between virial temperatures estimated from the  
 velocity dispersion ${\sigma}$ (CNOC Survey C97) 
with measured temperatures : $^{a}$M\&S97; $^{b}$Donahue 1997; $^{c}$Tsuru 
et al. 1997; $^{d}$Henry 1997 .}  
\end{center}
\end{table} 

\subsection{ Deriving the temperature } 
 
As explained in section 3, our sample is not based on clusters with 
measured temperature only: 
it also contains distant clusters for which the temperature information has  
been estimated  in an indirect way.  
From the lensing analysis, S97 have used deep WFPC-2 imaging of 12 distant 
clusters at redshifts 
between  
$z = 0.17$ and 0.56. Using the distortion  of faint galaxies detected in 
these fields, they 
measured  the mean shear 
and inferred a mass estimate within a  
radius of $200 h^{-1}$ kpc from the cluster lens center, assuming 
a singular isothermal profile $\propto r^{-2}$. 
We have used these masses to derive the X-ray 
temperature of these 11 
distant clusters,   
applying the following scaling 
relation:\\ 
\begin{equation} 
T_{X} = 5.38 M_{lens} {\rm keV}  
\end{equation} 
where $ M_{lens}$ is in units of $10^{14}h^{-1}{\rm M}_{\odot}$.
For the C97 clusters, we have converted the virial masses derived 
inside the total radial extent of the sample  
from the measured velocity dispersions  
 to an X-ray temperature. For this, we used the scaling relations derived from  
Evrard's numerical simulations (Evrard 1989, Evrard 1997), 
we infer a relation between velocity dispersion and temperature: 
\begin{equation} 
T_{X}=({\sigma}/350{\rm km/s})^{2} {\rm keV} 
\end{equation} 
In some cases, it has been possible to compare the temperature as estimated 
from (3) and (4) with satellite temperature measurements (see Tables 3 and 4).
  
Although the number of clusters for which the information is available is 
small, the agreement is rather good (Figure 2). This suggests that lensing 
and the virial mass estimates are essentially in agreement with the X-ray mass 
at a level better than a factor of two, although one can notice a slight 
tendency towards overestimation, but the samples are too 
small to draw any firm conclusion. 
\begin{table}[t] 
\begin{center} 
\begin{tabular}{lllrr}
 \hline

Cluster & z &  $M_{lens}$&   $T_{X}^{lens}$&  $T_{X}$\\
        &   &  $10^{14}{\sf M_{\odot}}$& {\sf keV} & {\sf keV}\\  
\hline
 
A2218  & 0.17  & 1.05  &     5.65 &     $6.35^{a}$ \\ 
AC118  & 0.31  & 1.85  &     9.92 &     $9.95^{b}$ \\ 
CL0016 & 0.55  & 1.87  &    10.06 &     $7.55^{c}$\\ 
CL0939 & 0.41  & 0.73  &     3.93 &     $2.90^{d}$ \\ 
3C295  & 0.46  & 2.35  &    12.64 &     $7.50^{a}$\\
\hline
\end{tabular} 
\caption {\small Comparison between estimated temperatures from $ M_{lens}$ 
(weak lensing, S97) with measured temperatures: 
$^{a}$M\&S97; $^{b}$Tsuru et al. 1997; $^{c}$Hughes \& Birkinshaw, 1997;
$^{d}$Schindler 1997.} 
\end{center} 
 
\end{table} 
 
 \begin{figure}[h] 
\epsfxsize=9cm 
\centerline{\epsfbox{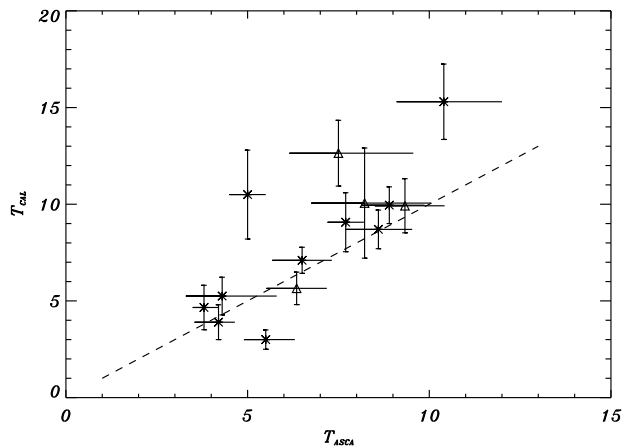}} 
\caption{\small \sl Comparison between estimated temperatures from $ M_{lens}$ 
 as inferred from weak lensing S97 (triangles) and from CNOC survey velocity dispersions (crosses) with measured temperatures $T_{X}$.}
\label{Fig. 2b} 
\end{figure} 
\noindent 
\section {Method and Results} 
\subsection{Investigating the possible evolution} 
 
\begin{figure*} 
\centerline{\psfig {figure=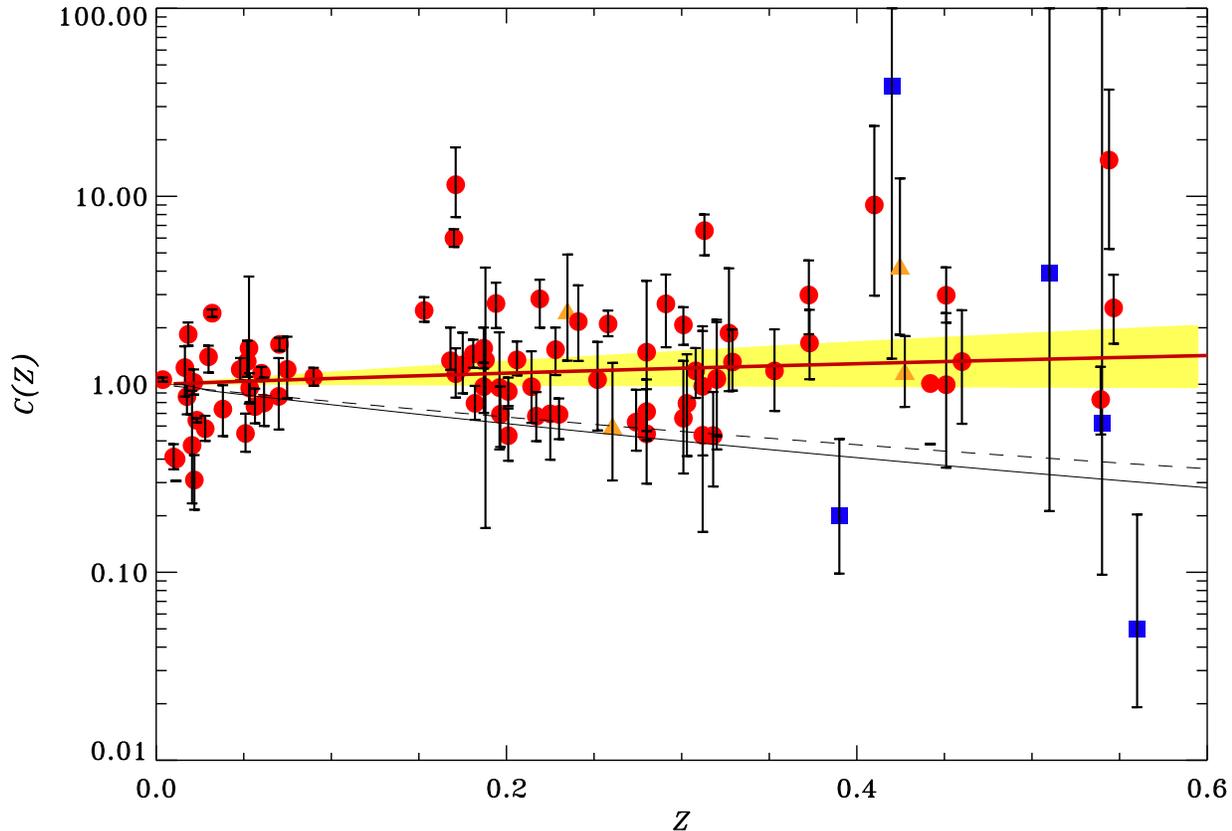,height=12cm}}
\caption{\small In this figure we present the 
coefficient $C_i$ for all the 
clusters in our sample with redshift smaller than 0.6. The filled circles are 
used for actual X-ray 
measurement of the temperature, filled squares are for weak lensing 
measurements of S97 and filled triangles are for the CNOC clusters. 
The error bars are   
$1 \sigma$. The thick line represents the best 
fitting power--law, 
the shaded area represents an estimate of the 90\% confidence range. 
 The prediction of
a low density universe ($\Omega_0 = 0.2$) is represented by the thin line.
The dashed, thin line is the curve when the correction in the luminosity 
distance is not taken into account in the expression of $C(z)$.}
\label{fig:cz} 
\end{figure*}

Ideally, one would like to estimate directly the $L_{X}-T_{X}$ in a redshift 
bin centered on a 
value of the redshift as high as possible. However, since the number of clusters 
decreases 
rapidly for redshifts greater than 0.25, it becomes difficult 
to get any information
from this method at high redshift:  dividing the  
sample in several redshift bins and trying
 to fit the luminosity--temperature relation 
in the different redshift bins becomes unpractical
for redshifts greater than 0.35. 
Another method has been used which consists in plotting 
the mean temperature of clusters above some threshold luminosity in order 
to minimize any 
possible systematic effect (Arnaud et al, 1991, M\&S97). 
It remains possible, however, 
that higher redshift clusters are brighter in the mean, introducing a bias 
in the sample.
Furthermore, this method results in a rude elimination of some of the data.  
We have tried to find an efficient estimator of the 
evolution which is adequate for the kind of evolution introduced by OB97. 
To get  round this problem we have introduced a new evolution estimator. 
For each cluster $i$,with measured $L_i, T_i,$ we have estimated the  
following quantity: 
\begin{equation} 
C_i =  \frac{L_i}{L_1T_i^{\alpha}} 
\label{eq:cz}
\end{equation} 
where   $ L_1 = 0.049 {10^{44} {\rm erg/s/cm}^2}$ and $\alpha = 3$. 
The dependance of $C_i$ on redshift probes the evolution:
clearly, if the cluster population is not evolving, the mean value of 
this quantity should 
remain constant with redshift. 
Note that it is possible that the  $L_{X}-T_{X}$ evolves and that the 
measured $C$ will not probe this evolution, this would need, however, some kind
of conspiracy. 
Because $L_X$ is estimated from the apparent flux, $C(z)$ also contains
a term coming from the cosmological parameters of the universe in the 
luminosity distance. In practice, $L_X$ is estimated in a Einstein-de 
Sitter universe; therefore,  
the theoretical value $C(z)$ has to be corrected when one is comparing
data with low-density universe predictions  (however, this term is small, 
as can be 
seen in  fig. \ref{fig:cz}).
It is also clear that this method can be applied without removing any 
data, and that it takes fully taking into account the information in redshift.
We have applied this test to our sample using the OB97 parameters for the 
$L_{X}-T_{X}$ relation. For each cluster, we have computed 
the quantity given by (5) and estimated an uncertainty range 
on this quantity, neglecting 
the uncertainty in the 
luminosity. The result is plotted in Fig \ref{fig:cz}.

The measure of evolution can now be directly obtained by fitting 
a power law to the data:
\begin{equation}
  C(z) = \alpha(1+z)^{\beta}
\end{equation}
in which $\alpha$ and $\beta$ are determined by a likelihood analysis.
However, the intrinsic 
dispersion of the values
of $C$ is often higher than the uncertainty on the coefficient 
itself, due to the errors in the 
temperature. 
We have therefore estimated the intrinsic dispersion of $C$ from
our template sample and added it 
in quadrature with the uncertainty on the
individual values of the coefficient. We have also checked that 
the results are insensitive to the assumed dispersion. 
In order to investigate the robustness of our results, 
we have performed 
various analyses, whose results are summarized in Table 5:
for various sub-samples, we have reported the best estimated parameters from 
the likelihood analysis and the 68\% and 90\% confidence limits.
The expected value of $\alpha$ is 1. If only 
clusters with redshift greater than 0.15 are used, there is a degeneracy 
between $\alpha$ and $\beta$: in the two-parameter plane, 
the confidence domain 
looks like an elongated ellipse, allowing a wide range of $\beta$, but for 
unrealistic values of  $\alpha$. This is the main motivation to include a 
set of low 
redshift clusters in our sample, before performing the likelihood analysis.
With this approach, $\alpha = 1$ value always falls in the 90\% confidence range
when $\alpha$ and $\beta$ are determined by a likelihood analysis.
In our analysis, we have generally assumed a gaussian distribution of 
the errors, which is
equivalent to a $\chi^2$ minimization. When the full sample is used, 
the $\chi^2$ of the best fitting model is rather poor : $\chi^2$ is equal 
to 219 for 97 clusters. This is due to a few outliers, which are far away from 
the general trend. This can be seen by using the $l_1$ norm instead of the 
standard  $\chi^2$: five clusters were found to have high $C$ value
(A1204,  RXJ 1340+4018, MS2137, Cl 0939+47, RXJ 0018+16) and were removed 
in most sub-samples (in this case the sub-sample is flagged by -o in 
Table 5).\\ 
 \begin{table*}
\begin{center}
\begin{tabular}{lrrllc}
  \hline
\vspace*{-2mm}\\
Sub-sample      & $N _c$ &    $\delta \chi^2$  &        $\alpha\pm 68\% \pm 90\%$           & $\beta\pm 68\% \pm 90\%$    & $\Omega_0\pm 90\%$ \vspace*{1mm}\\
\hline 
\vspace*{-2mm}\\
Complete sample & 71 & 193. & $1.15^{+0.10+0.20}_{-0.10-0.15}$ & \hspace*{2.5mm}$ 0.00^{+0.30+0.50}_{-0.30-0.60}$ & $0.75^{+0.12}_{-0.15}$\vspace*{1mm}\\
($L_1$ norm) & 71 & 63. & $1.05^{+0.15+0.25}_{-0.15-0.25}$ & $ -0.10^{+0.60+1.00}_{-0.60-1.10}$ & $0.72^{+0.25}_{-0.27}$\vspace*{1mm}\\
X-ray sample & 57 & 182. & $1.14^{+0.10+0.15}_{-0.10-0.15}$ & $ +0.28^{+0.30+0.50}_{-0.35-0.60}$ & $0.82^{+0.15}_{-0.17}$\vspace*{1mm}\\
           & 57 & 184. & {\bf 1.00 }                    & $ +0.65^{+0.25+0.35}_{-0.20-0.35}$ & $0.91^{+0.09}_{-0.09}$\vspace*{1mm}\\
X-ray sample -o & 52 & 68. & $1.07^{+0.10+0.17}_{-0.05-0.13}$ & $+0.20^{+0.30+0.60}_{-0.35-0.70}$  & $0.80^{+0.15}_{-0.17}$\vspace*{1mm}\\
           & 52 & 69. & {\bf 1.00 }                    & $+0.40^{+0.20+0.40}_{-0.30-0.50}$ & $0.85^{+0.10}_{-0.12}$\vspace*{1mm}\\
X-ray sample $0.15<z<0.6$ -o & 49 & 62. & $0.98^{+0.10+0.17}_{-0.10-0.15}$ & $+0.95^{+0.35+0.65}_{-0.40-0.75}$  & $0.99^{+0.16}_{-0.19}$\vspace*{1mm}\\
                             & 49 & 62. & {\bf 1.00 }                    & $+0.82^{+0.35+0.45}_{-0.25-0.50}$ & $0.95^{+0.11}_{-0.12}$\vspace*{1mm}\\
X-ray sample $0.15<z<0.35$ -o & 41& 49. & $1.00^{+0.12+0.18}_{-0.10-0.18}$ & $ +0.65^{+0.50+1.00}_{-0.65-1.05}$  & $0.92^{+0.25}_{-0.26}$\vspace*{1mm}\\
                    & 41 & 49. & {\bf 1.00 }   & $ +0.65^{+0.35+0.55}_{-0.35-0.60}$ & $0.92^{+0.14}_{-0.15}$\vspace*{1mm}\\
X-ray sample $0.35<z<0.6$ -o & 8  & 11. & $0.95^{+0.10+0.17}_{-0.10-0.17}$ & $+1.30^{+0.45+0.75}_{-0.60-1.05}$  & $1.07^{+0.20}_{-0.25}$\vspace*{1mm}\\
                            & 8 &  11. & {\bf 1.00 }                    & $+1.10^{+0.40+0.60}_{-0.55-0.80}$ & $1.02^{+0.15}_{-0.20}$\vspace*{1mm}\\
EMSS & 16 & 52. & $1.00^{+0.12+0.20}_{-0.08-0.15}$ &  $ +0.00^{+0.50+0.80}_{-0.60-1.10}$  & $0.75^{+0.20}_{-0.27}$\vspace*{1mm}\\
     & 16 & 52. & {\bf 1.00 }    & $ +0.10^{+0.30+0.60}_{-0.40-0.80}$ & $0.77^{+0.15}_{-0.20}$\vspace*{1mm}\\
EMSS $z<0.6$ -o & 14 & 11.6 & $0.95^{+0.12+0.20}_{-0.10-0.15}$ & $ +0.25^{+0.75+1.00}_{-0.75-1.35}$  & $0.81^{+0.25}_{-0.34}$\vspace*{1mm}\\
         & 14 & 11.7 & {\bf 1.00 }                    & $ +0.15^{+0.50+0.80}_{-0.55-1.05}$ & $0.79^{+0.20}_{-0.26}$\vspace*{1mm}\\
Henry -o & 9 & 7.5 & $0.95^{+0.10+0.20}_{-0.10-0.17}$ & $ +0.60^{+0.80+1.40}_{-1.00-1.75}$  & $0.90^{+0.35}_{-0.40}$\vspace*{1mm}\\
          & 9 & 7.6 & {\bf 1.00 }                    & $ +0.40^{+0.60+0.95}_{-0.70-1.20}$ & $0.85^{+0.24}_{-0.30}$\vspace*{1mm}\\
 CNOC  ($T_{est}$) & 15 & 2. & $1.05^{+0.10+0.20}_{-0.15-0.20}$ & $-1.50^{+1.30+2.10}_{-1.50-3.30}$  & $0.37^{+0.50}_{-0.82}$\vspace*{1mm}\\
X-ray CNOC ($T_X$)  & 10 & 18. & $0.95^{+0.10+0.20}_{-0.10-0.15}$ & $ +1.20^{+0.50+0.90}_{-0.70-1.20}$ & $1.05^{+0.22}_{-0.30}$\vspace*{1mm}\\
Smail sample ($T_{est}$)& 10 & 3.3 & $1.05^{+0.15+0.25}_{-0.10-0.20}$ & $-2.00^{+1.00+1.50}_{-1.70-3.00}$  & $0.25^{+0.40}_{-0.75}$\vspace*{1mm}\\
X-ray Smail sample ($T_X$) & 5 & 4. & $0.90^{+0.15+0.20}_{-0.10-0.15}$ & $ +1.90^{+0.60+0.90}_{-0.90-1.50}$ & $1.23^{+0.22}_{-0.37}$\vspace*{1mm}\\
X-ray Smail sample ($T_{est}$) & 5 & 2.& $1.05^{+0.15+0.20}_{-0.10-0.15}$ & $ +0.90^{+1.30+1.90}_{-1.60-2.60}$ & $0.52^{+0.48}_{-0.65}$\vspace*{1mm}\\
    \hline
        \end{tabular}
 \caption{This table summarises the main results of the various fits we have performed. 
The full sample with 57 X-ray clusters with measured X-ray temperatures plus 4 
CNOC clusters and 5 S97 clusters for which the temperature is derived from the virial velocity dispersion and lensing mass respectively. 
Col. 2 gives the number of high redshift clusters ($z>0.15$), the $\chi^2$ value in col. 3 ( the contribution of the low redshift 
cluster subset to the $\chi^2$ is nearly 
constant and equal to 27.), the best fit estimates of $\alpha$ and 
$\beta$ with the uncertainties at  
68\% and 90\% confidence level are given in col. 4 and col. 5 respectively 
and the corresponding value of $\Omega_0$ as inferred from eq. 7 
and the 90\% uncertainties are given in col. 6.
}
   \label{tabres}
\end{center}
            \end{table*}
The X-ray data, even when they are split into  various sub-samples  
(like the EMSS X-ray 
clusters alone or the Henry set alone) always lead to a high value of the 
best $\beta$, in the range $[0.-1.0]$. It is important to note that no 
systematic trend is found: the various X-ray selected samples always give a 
$\beta$ in the same range, with an uncertainty at worst of the order of 0.5.   
Therefore the range $[0.-1.0]$ can be taken as the 90\% 
confidence interval. The only two samples 
which do not lead to a high value of $\beta$ are the CNOC sample and 
the S97 sample with their
estimated temperatures (from virial and lensing mass estimates). However, 
when these samples are 
restricted to clusters for which the temperature has been measured 
(the X-ray CNOC and X-ray S97 sample in Table 
3), the resulting $\beta$ is much higher and in agreement with other 
analysis based on X-ray data. This may be due to the fact that lensing 
and virial mass estimates are higher than X-ray masses, which introduces a 
bias toward lower $C$ and, consequently, to artificially lower $\beta$ at 
high $z$. \\
Our result confirms previous investigations: available data on X-ray clusters
are consistent with the lack of significant evolution in the 
$L_X-T_X$ relation. Finally, 
it is somewhat troublesome 
that the  highest redshift clusters ($Z > 0.8$) lead to a small value of 
$C$. Better temperature and flux 
measurements 
will be of great interest, and we obviously need more (very) 
high redshift clusters temperature measurements.\\ 

\begin{figure} 
\vspace*{-1.5cm}
\hspace*{1cm}
\centerline{\psfig {figure=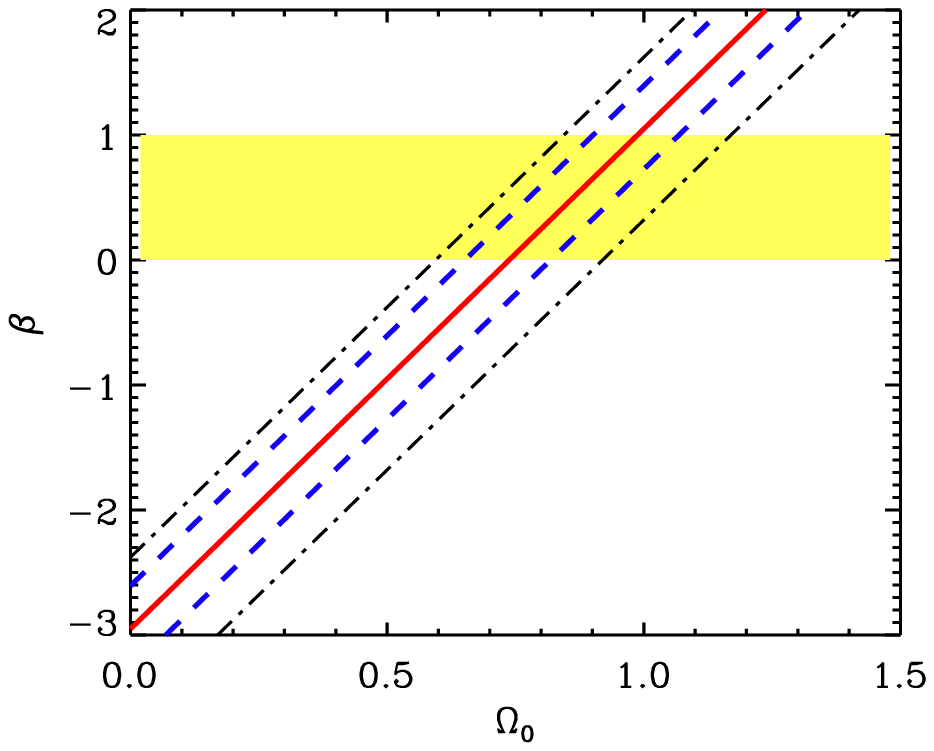,height=11cm}}
\vspace*{-2cm}

\caption{\small \sl Comparison of the predicted values of the coefficient 
to the values 
we have estimated from the observations. The grey area corresponds to 
our 90\% confidence range for 
$\beta$. The solid thick line corresponds to the best 
evolution law  when the EMSS redshift distribution is fitted, the dashed 
(resp. 
dotted-dashed) correspond to an estimation of the 68\%  contour (95\% resp.). } 
\label{fig.om} 
\end{figure} 

\subsection{What do the data tell us?} 
 
As discussed previously, the knowledge of the evolution of the number density 
of clusters with redshift is a powerful cosmological test. By fitting the 
EMSS cluster redshift distribution, OB97 showed that the knowledge of 
evolution of the temperature-luminosity relation 
provides an alternative way to measure 
the mean density of the universe.
 We have iterated OB97 analysis 
for various values of the density parameter: the temperature 
distribution $N(T_x)$  was fitted in order to derive the power spectrum 
index of the fluctuations as well as its normalization; the best $\beta$ in 
eq. (\ref{LTevol}) was then determined by fitting  the  
EMSS redshift distribution, as well as the 1- and 2- sigma interval. The best 
fitting parameter $\beta$ is tighly related to $\Omega_0$ accordingly to 
the following relation:
\begin{equation}
\beta = 4.\times \Omega_0 -3.  
\end{equation}
This relation, as well as the 1- and 2- sigma interval contours,
 are presented on figure \ref{fig.om}.  The high value of the slope 
illustrates the strong dependence 
on ${\Omega_0}$, highlighting the power of this test.\\

From the range of $\beta$ we estimated from the data, we can obtain 
an estimation of $\Omega_0$. 
This is presented in figure \ref{fig.om}, where we have plotted the 
value of $\beta$
which is necessary to fit the EMSS cluster redshift distribution as 
well as the confidence range. The grey area represents the range $[0.-1.]$, 
which is used as our 90\% range obtained from the likelihood analysis. 
As one can see, the range of values we obtained favors a high 
density universe with a formal determination of the mean density parameter:
$\Omega_0 = 0.85^{+0.2}_{-0.2}$.  The main potential problem in our analysis 
is that the correspondence 
between $\beta$ and $\Omega_0$ has been obtained by OB97 from the EMSS survey,
while the sample we used consists of clusters for which selection rules 
cannot be defined in a simple way. There are some limitations in 
the present 
work which imply that the conclusion should be considered as only 
preliminary: firstly, because the 
selection function of X-ray clusters in EMSS is not well understood,
it conceivable that a substantial number of clusters have been missed
(for instance because their surface brightness was too low). This is not 
supported by the fact that the modeling of X-ray clusters as done by OB97
predicted correctly the abundance of faint clusters as detected by ROSAT.
Secondly, it is possible that our sample contains clusters 
which have preferentially high $\beta$ as we have already mentioned;
 it is possible that these clusters
suffer from a systematic bias favoring high $\beta$. Still, we do not find 
any evidence of such a bias. For instance, the typical 
luminosity in the sample 
does not seem to increase with redshift. In general, no systematic tendency 
was found 
by eye inspection.  Thirdly, OB97  normalized the models at $z = 0$ by use of 
the HA91 temperature distribution function, which may suffer from systematic 
uncertainties (Eke at al., 1996, Henry, 1997). 
A more detailed  investigation of these various  
questions would need a  Monte-Carlo simulation as well as 
a systematic investigation of possible biases. This will be addressed in a 
future paper. 
 
\section{Conclusion and discussion}

In this paper, we have addressed the question of the cosmological evolution 
of the $L_{X}-T_{X}$ relation and we have investigated the possible  
cosmological implications of this evolution (or lack of it). 
The new indicator $C(z)$ is well adapted to measure the 
possible evolution of the $L_{X}-T_{X}$ relation when the number of 
available clusters is small.  We have applied this measure to a sample 
of high and intermediate redshift clusters for which the temperature 
information is available. We have found no strong evidence 
of evolution 
of the $L_{X}-T_{X}$ relation, in agreement with other works. Despite the 
fact that our sample is not complete and that the number of 
high redshift clusters is not large, we have shown that our method 
is robust and leads to statistically significant result. 
We have furthermore used our results on  $L_{X}-T_{X}$ evolution 
to constrain the value of ${\Omega_o}$ accordingly to the test of OB97. 
This is indeed the first 
time that this  test is applied to 
observations. This test is extremely powerful because it results from a 
fundamental difference between high-- and low--density universes:
the rate of structure formation. Therefore, it provides a
global test of the mean density of the universe, rather than a local 
dynamical one, as are classical $M/L$ estimates. One can therefore expect to
obtain in this way a definitive answer on the value of the mean density 
of the universe.

The absence of negative evolution in the $L_{X}-T_{X}$ relation, as we 
have found, 
provides an indication of a high-density universe: accordingly to our analysis,
the range of evolution we find is consistent with  
$\Omega_0 = 0.85^{+0.2}_{-0.2}$ (at 90\% confidence level). The fact 
that our sample is not drawn 
from an X-ray selected sample implies a possible bias and 
therefore our conclusion can be considered 
as only preliminary. A more robust answer on the mean density of the universe 
can be obtained only from a well-controlled X-ray selected sample of clusters.
Our work shows that even if no definitive 
conclusion can be drawn, a reasonable number of X-ray temperature measurements
can provide a very interesting answer on the mean density of the universe
and that open model universes seem to be facing a 
serious problem. It is interesting to note that Barbosa et al. (1996) 
have also pointed out 
an other piece of evidence coming from clusters which 
disfavors open model universes.  
It is realistic to envisage that a definitive answer could be obtained
with XMM by a follow-up of a sample of high redshift clusters selected from
an X-ray flux limited survey.
  
\begin{acknowledgements}We acknowledge useful discussions with M.Arnaud 
and J.Bartlett,as well as  the referee, G.Evrard, whose comments help to 
clarify this paper.
\end{acknowledgements}

\end{document}